\documentclass[sigconf]{acmart}
 \renewcommand\footnotetextcopyrightpermission[1]{} 
 \pdfoutput=1
\usepackage{tikz}
\usepackage{multirow}
\usepackage{amsmath}
\usepackage{qcircuit}
\usepackage{algorithm}
\usepackage{algpseudocode}
\usepackage{graphicx}
\usepackage{caption}
\usepackage{standalone}
\usepackage{subcaption}
\usepackage{float}

\newcommand{\name}{EPOC}

\usetikzlibrary{shapes,arrows,positioning}
\usetikzlibrary{shadows,arrows.meta,positioning,backgrounds,fit}
\tikzset{%
  materia/.style={draw, fill=blue!20, text width=6.0em, text centered, minimum height=1em,drop shadow},
  etape/.style={materia, text width=18em, minimum width=21em, minimum height=2em, rounded corners, drop shadow},
  etape2/.style={materia, fill = white, text width=14em, minimum width=14em, minimum height=2em, rounded corners},
  texto/.style={above, text width=6em, text centered},
  linepart/.style={draw, thick, color=black!50, -LaTeX, dashed},
  line/.style={draw, thick, color=black!50, -LaTeX},
  ur/.style={ text centered, minimum height=0.01em},
  back group/.style={fill=yellow!20,rounded corners, draw=black!50, dashed, inner xsep=10pt, inner ysep=10pt},
  back group2/.style={fill=white,rounded corners, draw=black!100, dashed, inner xsep=10pt, inner ysep=10pt},
}

\usetikzlibrary{external}
\begin{document}

\title{EPOC: An Efficient Pulse Generation Framework with Advanced Synthesis for Quantum Circuits }

\author{Jinglei Cheng}
\authornote{Both authors contributed equally to this research.}
\affiliation{%
  \institution{Purdue University}
  \city{West Lafayette}
  \state{IN}
  \country{USA}
}
\author{Yuchen Zhu}
\authornotemark[1]
\affiliation{%
  \institution{Rensselaer Polytechnic Institute}
  \city{Troy}
  \state{NY}
  \country{USA}
}
\author{Yidong Zhou}
\affiliation{%
  \institution{Rensselaer Polytechnic Institute}
  \city{Troy}
  \state{NY}
  \country{USA}
}

\author{Hang Ren}
\affiliation{%
  \institution{University of California, Berkeley}
  \city{Berkeley}
  \state{CA}
  \country{USA}}
\author{Zhixin Song}
\affiliation{%
  \institution{Georgia Institute of Technology}
  \city{Atlanta}
  \state{GA}
  \country{USA}}
\author{Zhiding Liang}
\affiliation{%
  \institution{Rensselaer Polytechnic Institute}
  \city{Troy}
  \state{NY}
  \country{USA}
}


  

  
  



\begin{abstract}

Quantum optimal control has been explored by researchers due to its capability to greatly reduce circuit latency. 
However, it is also known for its significant computational overhead. 
Previous works have proposed various methods to accelerate quantum optimal control, such as utilizing GPUs, pre-compilation techniques, and improved circuit libraries. 
These pulse generation frameworks focus on generating pulses from unitary matrices derived from quantum circuits, without exploring better unitary matrices through equivalent representations. 
Consequently, they overlook many optimization opportunities by adopting coarse-grained methods.
In this work, we propose a novel approach that combines ZX-Calculus, circuit partitioning and circuit synthesis to accelerate pulse generation. 
Our contribution lies in employing finer granularity in pulse generation, enabling increased parallelism and decreased latency in quantum pulses. 
Finer granularity is achieved by grouping quantum gates and decomposing the resulting unitary matrices into smaller unitary matrices using synthesis techniques.
Additionally, we explore further optimization possibilities by continuously optimizing the circuit through the identification of equivalent representations. 
By adopting these techniques, we achieve a further reduction in circuit latency while only requiring quantum optimal control for relatively small-sized unitary matrices. 
For the first time,  circuit synthesis is introduced into the workflow of quantum optimal control. We are able to achieve \textbf{31.74\%} reduction in latency compared to previous work and a \textbf{76.80\%} reduction compared with the gate-based method to create pulses.
Our approach demonstrates the potentials for significant performance improvements in quantum circuits while minimizing the computational overhead associated with quantum optimal control.

\end{abstract}

\maketitle
\pagestyle{plain}
\section{Introduction}
\label{introduction}
Quantum computing has emerged as a promising paradigm with the potential to outperform classical computers in various domains~\cite{zhuang2022defending, jiang2021co}, including chemistry~\cite{levine2009quantum}, optimization~\cite{he2023alignment}, machine learning~\cite{liang2021can}, and physical simulations~\cite{buluta2009quantum}. 
However, the current Noisy Intermediate-Scale Quantum (NISQ) era~\cite{Preskill2018NISQ} presents significant challenges, such as limited qubit count, short coherence times, and high error rates~\cite{liang2024napa}. 
To exploit the power of NISQ devices effectively, it is crucial to develop efficient quantum circuit optimization techniques that mitigate the impact of noise and errors~\cite{cheng2023fidelity}. 
The coherence time determines the duration and depth of quantum circuits that can be successfully executed on real quantum hardware. 
The term ``quantum volume'' has been proposed to quantitatively describe this property, which takes into account both the number of qubits and the depth of the quantum circuit that can be reliably executed on a given quantum device~\cite{gokhale2020optimization}. 

\begin{figure}
    \centering
    \includegraphics[width=\linewidth]{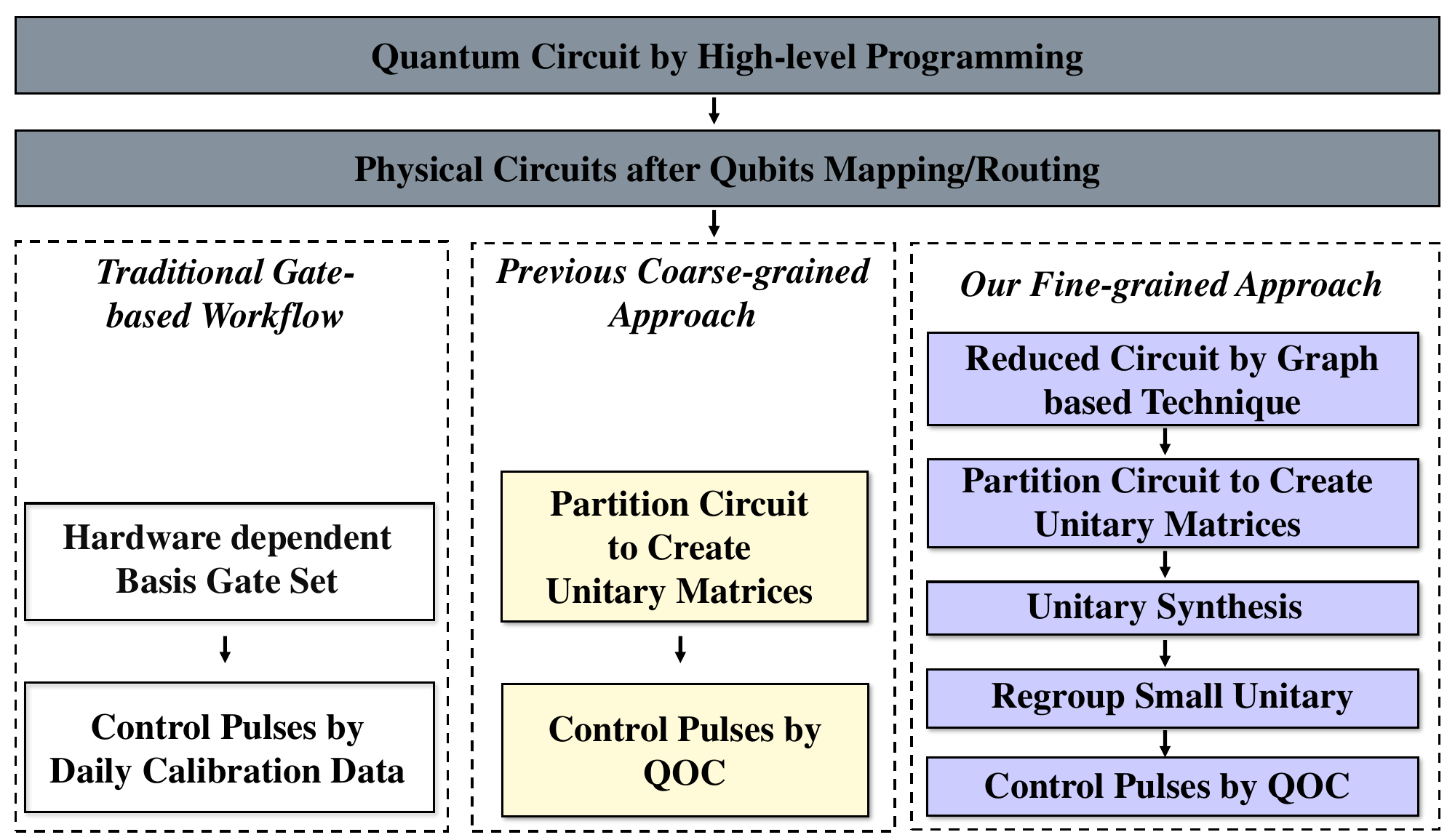}
    \caption{The diagram compares the traditional gate-based workflow for creating quantum circuits with our novel fine-grained approach. Our method introduces additional compilation steps, such as graph-based circuit reduction, circuit partitioning, and small unitary regrouping, to leverage the finer granularity provided by synthesis techniques.}
    \label{teaser}

\vspace{-3mm}
\end{figure}

A typical compilation workflow for quantum programs begins with the implementation of quantum algorithms by quantum circuits. 
As shown in the Figure~\ref{teaser}, these quantum circuits undergo different compilation passes to become compatible with the underlying hardware topology. 
For example, the circuit must be decomposed into basis gates and mapped according to the target quantum computer's architecture~\cite{li2019tackling}. 
Once the quantum circuits are expressed as basis gates, these gates are converted into physical signals that are applied to the qubits~\cite{krantz2019quantum}.
In the case of superconducting quantum computers, the physical signals are modulated microwave pulses that are sent to transmons, which are the basic building blocks of the superconducting quantum processor~\cite{liang2024napa}. 
Modulated microwave pulses is generated with an envelope that determines the function of the pulses and a carrier signal. 
\begin{figure}
\vspace{-3mm}
    \centering
    \includegraphics[width=\linewidth]{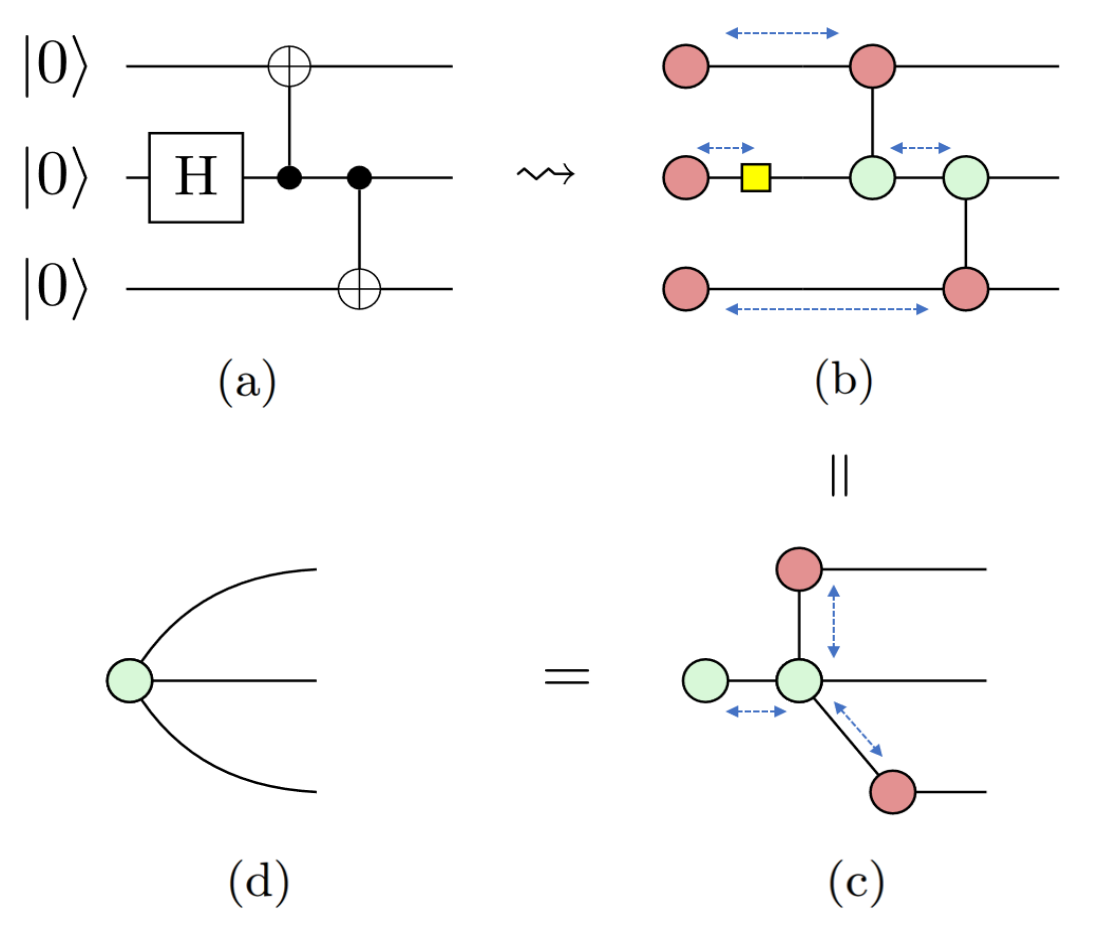}
    \caption{ZX-Diagram of a GHZ-state preparation circuit. (a) Quantum circuit for preparing a 3-qubit GHZ state $(|000\rangle+|111\rangle)/\sqrt{2}$. (b) Corresponding ZX-diagram. Simplifications are made using rewriting rules such as spider fusion, which is indicated by dashed double-arrow lines for contractable spiders. (c) Further simplification. (d) Final compact representation of the GHZ state~\cite{wikizx}.}
    \label{fig:ghz_zx}
\vspace{-3mm}

\end{figure}

In the final step of the compilation process, where basis gates are translated to pulses, there is one technique known as quantum optimal control (QOC)~\cite{werschnik2007quantum}. 
By finding the optimal control pulses that implement the desired unitary operation, QOC can minimize the effects of noise and errors~\cite{koch2022quantum}. 
However, QOC comes with a significant computational overhead, limiting its scalability~\cite{bukov2018reinforcement}. 
Previous works have proposed various methods to accelerate QOC, such as utilizing GPUs, pre-compilation techniques, and improved circuit libraries~\cite{leung2017speedup,cheng2020accqoc,chen2023pulse}. 
These pulse generation frameworks focus on generating pulses from unitary matrices directly derived from quantum circuits without exploring better unitary matrices through equivalent representations, thus overlooking many optimization opportunities by adopting coarse-grained methods. 

By coarse-grained methods, we refer to existing frameworks such as AccQOC~\cite{cheng2020accqoc} and PAQOC~\cite{chen2023pulse}, which perform circuit partitioning at the gate level. 
These methods first group the quantum gates in the circuit and calculate the corresponding unitary matrices for each group. 
QOC is then applied to these unitary matrices to generate the optimized control pulses. 
While this approach can provide some benefits in terms of pulse optimization, it is limited by the inherent gate structure of the quantum circuit. 
If the depth of the circuit blocks does not match, leading to mismatched pulse latencies, the utilization rate of the qubit lines, where the pulses are applied, may not be optimal. 
This can result in suboptimal performance and longer overall execution times.

In this paper, we aim to propose a fine-grained approach in contrast to the previous coarse-grained methods and traditional workflow, as depicted in Figure \ref{teaser}. Thus, we introduce \name, a novel approach that combines ZX-calculus~\cite{van2020zx}, circuit partitioning, and circuit synthesis with QOC to enhance pulse generation. 
Our main contribution is the development of a finer-grained method for generating quantum pulses, which results in increased parallelism and reduced latency. 
This is accomplished by partitioning circuits into blocks, computing their corresponding unitary matrices, and applying synthesis techniques to identify equivalent circuits composed of unitary gates. 
By breaking down relatively large circuit blocks into smaller unitary gates, we create additional opportunities for optimization, enabling more efficient and effective pulse generation.
However, directly applying QOC to these small unitary gates, which are the products of synthesis, presents a challenge. 
The issue arises because these blocks are typically too small to fully leverage the advantages of QOC. 
To address this, we need to aggregate these blocks to achieve a suitable size that allows for effective application of quantum optimal control techniques.
Therefore, we introduce a regrouping process for the unitary gates, where different unitary gates and two-qubit gates like CNOT are aggregated into a single unitary matrix. 
This regrouping technique allows us to further trade classical computational power for quantum computational power while limiting the size of the resulting matrices to avoid excessive computational overhead~\cite{cheng2020accqoc}. 


Finally, we apply QOC to the regrouped unitary matrices to obtain the optimized microwave pulses for the qubits. 
By combining QOC with the synthesized and regrouped unitary matrices, we generate highly optimized and robust quantum circuits tailored to the specific characteristics of the target quantum hardware. 
Our approach demonstrates the potential for significant performance improvements in quantum circuits while minimizing the computational overhead associated with QOC.

The main contributions can be summarized as follows:

\begin{itemize}
\item We propose \name, a novel framework bridging ZX-calculus, circuit partitioning, and circuit synthesis with quantum optimal control to accelerate pulse generation. 
\item We introduce a regrouping process that aggregates unitary gates and two-qubit gates into larger unitary matrices, providing additional opportunities for optimization. 
\item We demonstrate the effectiveness of \name\  through simulations and benchmarking on various quantum circuits. Our results show significant improvements in circuit latency, parallelism, and overall performance compared to state-of-the-art methods.
\end{itemize}

The remainder of this paper is organized as follows. Section~\ref{introduction} provides an introduction, followed by the necessary background information in Section~\ref{background}. We introduce our proposed framework, \name, in Section~\ref{methodology}. The evaluation of \name\ is presented in Section~\ref{evaluation}. Section~\ref{relatedwork} discusses related work, and we conclude the paper in Section~\ref{conclusion}.

\begin{figure*}
\vspace{-1cm}
    \centering

   \includegraphics[width=\textwidth]{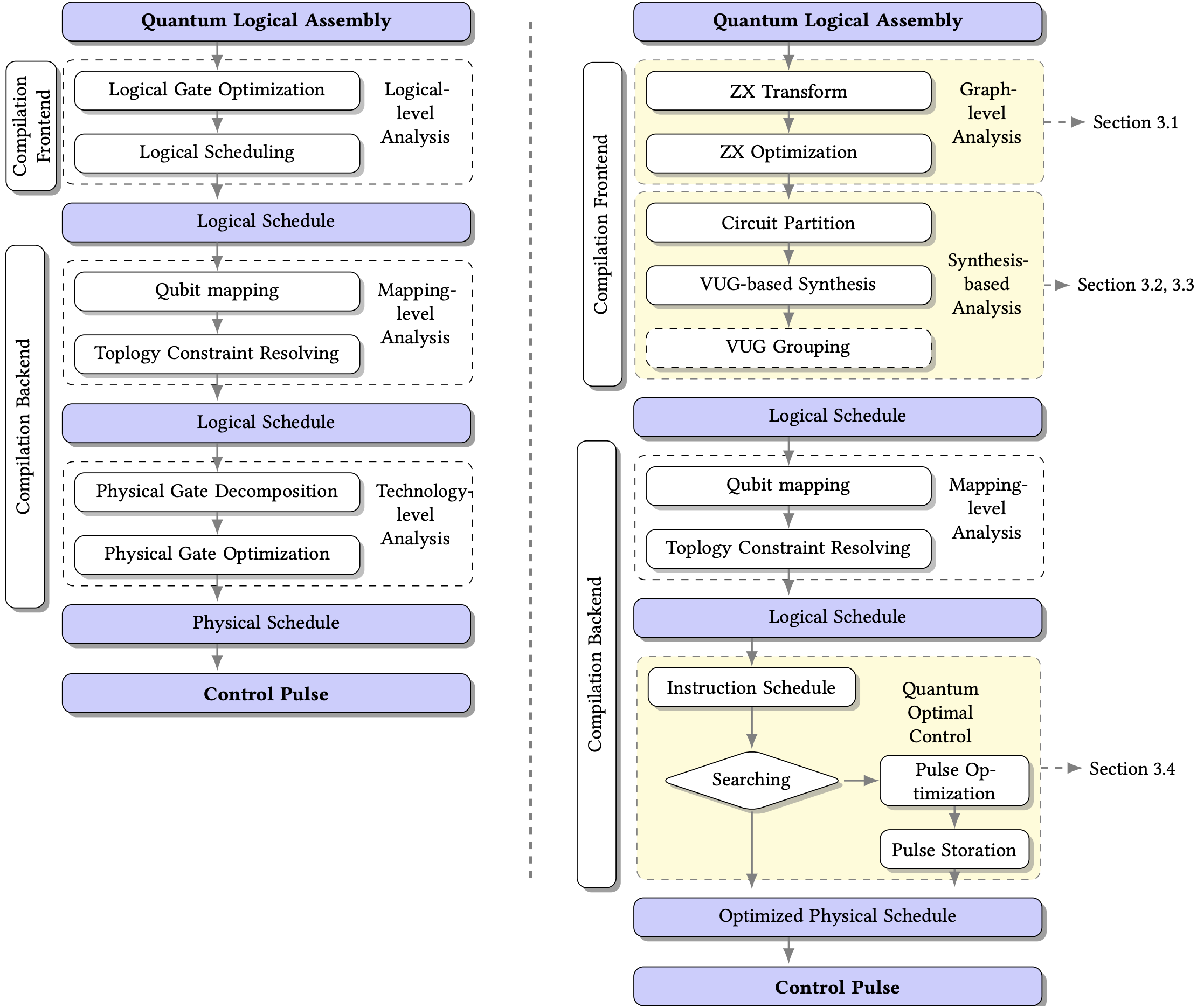}

    \caption{The comparison between standard gate-based compilation (left) and our approach (right). The colored areas highlight the major differences. In the first yellow box, our compiler converts input circuits into ZX graphs and performs ZX optimization to reduce the circuit depth. Then the optimized circuit is partitioned into groups with up to 8 qubits and each group is synthesized with VUGs as denoted in the second yellow box. Finally, in the third box, for each instruction, optimized pulses are generated and stored in the library for further usage.}
    \label{fig:workflow-label}

\vspace{-3mm}    
\end{figure*}

\section{Background}
\label{background}
\subsection{Quantum Computing}
Quantum computing fundamentally differs from classical computing due to the unique properties of qubits or qubits~\cite{nielsen2010}. 
Unlike classical bits, qubits can exist in superpositions of states, represented by:
$\psi = \alpha|0\rangle + \beta|1\rangle$
where $\alpha$ and $\beta$ are complex coefficients satisfying $|\alpha|^2 + |\beta|^2 = 1$. 
A quantum system with $N$ qubits can exist in up to $2^N$ states, requiring $2^N$ complex parameters for a complete description. 
This exponential scaling allows for efficient representation and processing of vast amounts of information.
Quantum entanglement occurs when the state of one qubit cannot be described independently of another qubit, enabling tasks such as quantum teleportation~\cite{pirandola2015advances} and superdense coding~\cite{wang2005quantum}.
Qubit measurement collapses the superposition to a basis state ($|0\rangle$ or $|1\rangle$) with probabilities derived from $|\alpha|^2$ and $|\beta|^2$. 
Quantum algorithms manipulate superpositions and entanglements before the final measurement.
Quantum gates, represented by unitary matrices, perform operations on qubits. Single-qubit gates include Pauli gates (X, Y, Z), Hadamard gate (H), and phase shift gates (S, T). 
Multi-qubit gates, like the controlled-NOT (CNOT) gate, create entanglement between qubits.
These unique properties enable quantum computers to tackle problems intractable for classical computers, particularly in cryptography, optimization, and quantum simulation.

\subsection{ZX-Calculus}

The ZX-calculus, developed by Bob Coecke and Ross Duncan \cite{10.1007/978-3-540-70583-3_25, Coecke_2011}, offers a graphical language tailored for representing quantum states and operations on qubits. 
It utilizes the formalism of string diagrams, with Pauli-X and Pauli-Z matrices as representing basis, thus termed ZX-diagrams.
ZX-diagrams represent the topological configuration of quantum circuits, allowing transformations that do not alter qubit connections due to their topological invariance. 
The calculus includes a set of graphical rewrite rules, collectively known as the ZX-calculus rules.
Fundamentally, ZX-diagrams are composed of building blocks called ``spiders'', which represent various quantum computing elements such as isometries, states, unitary operations, and measurements in specific bases. 
These diagrams feature nodes colored in green and red, where green nodes correspond to the computational basis states $|0\rangle$ and $|1\rangle$, and red nodes represent the Hadamard-transformed basis states $|+\rangle$ and $|-\rangle$.
Additionally, Hadamard nodes are depicted as yellow boxes. 
Connections in the diagram are made by wires, which can curve, cross, and contract. 
In Fig. \ref{fig:ghz_zx}, we give an example showing the ZX-diagram of a GHZ state which illustrates the ZX-Calculus and the rewriting rule for circuit simplification.

The theoretical foundation of ZX-Calculus is rooted in the field of dagger compact categories~\cite{selinger2007dagger}, a discussion of which is beyond the scope of this paper but can be explored in further readings~\cite{vandewetering2020zxcalculus}.
ZX-calculus has found application in various domains of quantum information and computation, including measurement-based quantum computation \cite{10.1007/978-3-642-14162-1_24}, quantum error correction \cite{de_Beaudrap_2020, Duncan_2014}, and quantum circuit optimization \cite{Fagan_2019}. 
The application of ZX-calculus to quantum circuit optimization is particularly relevant to ours.
\subsection{Synthesis}

Quantum circuit synthesis plays a crucial role in quantum computing by generating efficient and optimized quantum circuits from high-level mathematical descriptions of quantum algorithms. 
Synthesis techniques aim to decompose unitary matrices, representing quantum transformations, into a sequence of elementary quantum gates that can be directly executed on quantum hardware~\cite{shende2005synthesis}. 
The quality of synthesized circuits is typically evaluated based on the number of gates, particularly the count of expensive two-qubit gates like CNOT, as well as the overall circuit depth~\cite{tan2020optimal}. 
Shorter circuits with fewer gates are highly desirable, especially for NISQ devices characterized by limited coherence times and noisy operations. 
Effective synthesis methods can significantly reduce the computational overhead and improve the reliability of quantum algorithms on real hardware~\cite{younis2021qfast,smith2023leap}. 
Moreover, quantum circuit synthesis enables hardware design exploration, algorithm discovery, and circuit optimization.
Existing synthesis approaches can be broadly categorized into top-down methods, which employ rule-based decomposition techniques, and bottom-up methods, which utilize numerical optimization and search strategies to construct circuits incrementally.

\subsection{Quantum Optimal Control}
In the context of a closed quantum system, the dynamics can be described by a Hamiltonian, given as:
\begin{equation}
    H(t) = H_0 + \sum_{j=1}^n u_j(t) H_j,
\end{equation}
where \( H_0 \) represents the drift Hamiltonian that accounts for the intrinsic evolution of the system, and \( H_j \) are the control Hamiltonians modulated by time-dependent control signals \( u_j(t) \).
The temporal evolution of the quantum state \( |\psi\rangle \) in this system is governed by the Schrödinger equation:
\begin{equation}
     \frac{d}{dt}|\psi(t)\rangle = -iH(t)|\psi(t)\rangle,
\end{equation}
assuming the initial state of the system at time \( t=0 \) is \( |\psi(0)\rangle \).

QOC aims to determine the optimal control signals \( u_j(t) \) that transition the system from an initial quantum state to a desired target state. 
This control strategy is particularly critical in achieving precise state transformations while considering system constraints and minimizing operational errors. 
To optimize the control signals, QOC employs a cost function, typically defined in terms of the fidelity between the achieved unitary transformation induced by the applied control signals and a desired unitary goal.
This fidelity metric quantifies the accuracy of the state transformation.
Advanced algorithms such as Gradient Ascent Pulse Engineering (GRAPE)~\cite{khaneja2005optimal} and Chopped Random Basis (CRAB)~\cite{caneva2011chopped} are utilized to systematically solve the QOC problem. 
These methods iteratively adjust the control signals to maximize the fidelity, ensuring that the final quantum state closely approximates the target state.

\section{Methodology}
\label{methodology}

In this section, we present the methodologies employed in \name\ to accelerate pulse generation. As shown in Figure~\ref{fig:workflow-label}, \name\ consists of several steps, each of which will be discussed in detail throughout this section. We will provide a comprehensive explanation of the techniques and algorithms used in each stage of the \name\ process, highlighting their contributions to improving the efficiency and effectiveness of pulse generation.

\subsection{Graph-based Depth Optimization}
\label{graph}

\begin{figure}[htb]
\centering
\begin{subfigure}[b]{\linewidth}
\centering
\includegraphics[width=\linewidth]{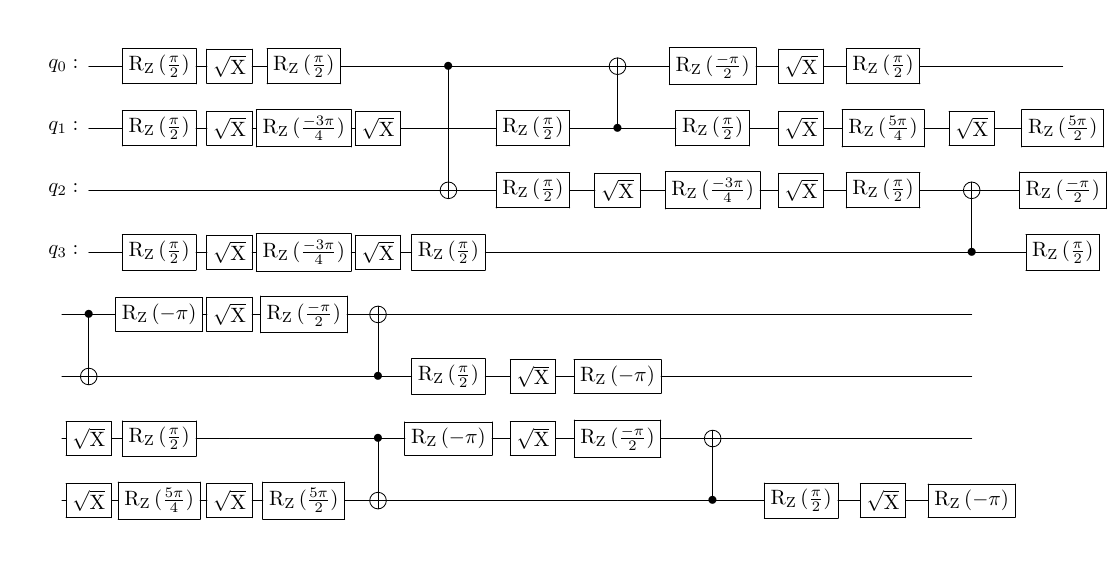}
\caption{Quantum Bell State Generation Circuit with 4 qubits. The circuit is implemented using basic quantum gates, including the rotation gate, SX gate, and CNOT gate. }
\label{circuita}
\end{subfigure}
\begin{subfigure}[b]{\linewidth}
\centering
\includegraphics[width=0.8\linewidth]{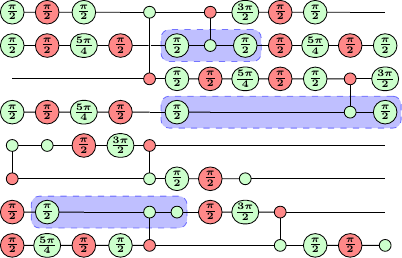}
\caption{ZX Graph Representation of the quantum circuit. The nodes within the blue box that share the same color are commutative.}
\label{circuitb}
\end{subfigure}
\begin{subfigure}[b]{\linewidth}
\centering
\includegraphics[width=\linewidth]{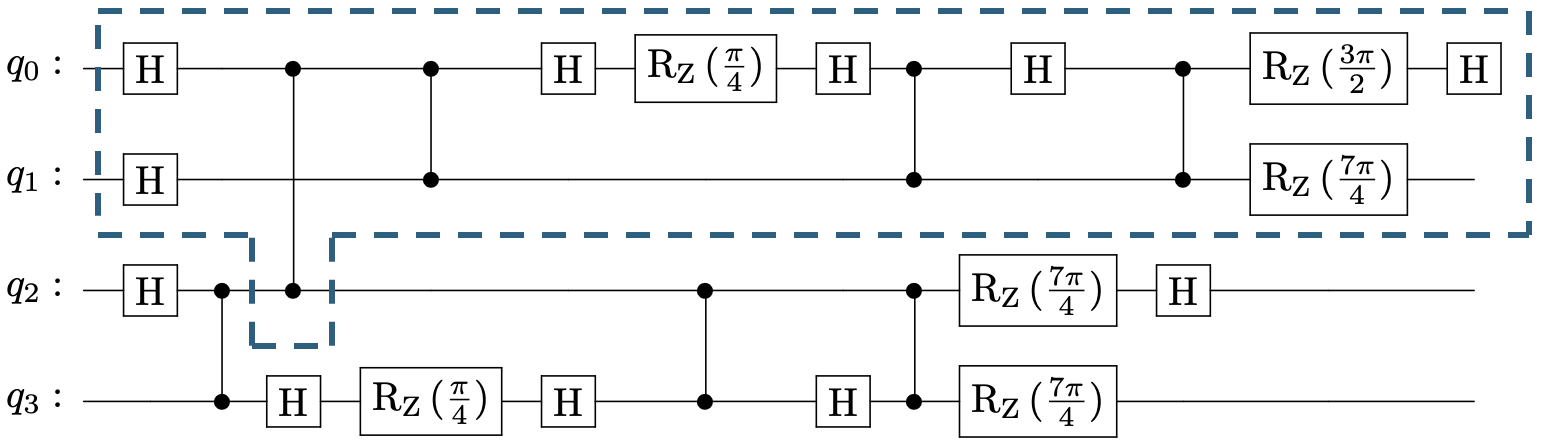}
\caption{Example of circuit grouping for the optimized circuit after applying ZX-calculus optimization.}
\label{circuitc}
\end{subfigure}

\caption{Graph-level optimization of the Bell state preparation quantum circuit (a) involves several steps: (b) transforming the circuit into a ZX graph representation, followed by the application of rewrite rules; (c) circuit grouping of the optimized circuit from the resulting ZX graph. }
\label{fig:zx}
\end{figure}
\begin{figure}[t]
\centering
\includegraphics[width=0.85\linewidth]{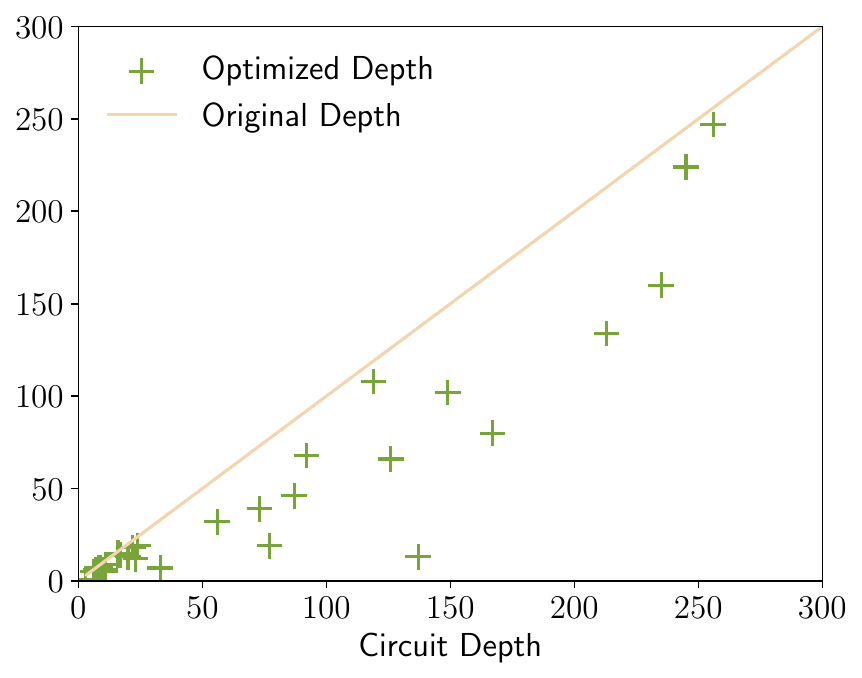}
\caption{ZX optimization results for 34 randomly selected quantum circuits. After applying ZX-calculus optimization techniques, an average depth reduction of 1.48 times is observed across the selected circuits.}
\label{zx_data}
\vspace{-3mm}
\end{figure}
After converting the input circuit into a ZX graph, the gates in the circuit are analyzed to determine if they can be delayed and potentially cancelled against future gates or combined with other gates. 
This process, known as gate commutation, allows for circuit optimization and depth reduction. 
For instance, a NOT gate can be placed on either side of a CZ gate for both control and target qubits without altering the gate's functionality. 
Similarly, a Z gate on the control qubit can be commuted through a CNOT gate if a Z gate also appears on the target qubit, and a NOT gate on the target qubit can be commuted through a CNOT gate if a NOT gate also appears on the control qubit. 
However, a Hadamard gate cannot be commuted through a CZ gate or a CNOT gate due to its non-commutative properties.
Gate commutation enables the aggregation of gates, leading to a reduction in circuit depth. 
For example, a NOT gate and a Hadamard gate can be combined into a Z gate, while a Z phase gate can be aggregated with a NOT gate, Z phase gate, or T gate by adjusting the rotation phase. 
Additionally, a pattern consisting of a Z phase gate between two Hadamard gates is transformed into a Hadamard gate between two Z phase gates with negative phases to enhance commutativity and aggregation capabilities.
Such a process is simplified with ZX graphs, as in a ZX graph, nodes in the same color are commutative and can be combined into a single node.
Apart from gate commutation and gate aggregation, other ZX-calculus rules such as spider fusion are applied to simplify the circuit representations and further reduce the complexity of the quantum circuit. 
In our work, we utilize PyZX ~\cite{kissinger2020pyzx}, a Python library designed for simplifying ZX-diagrams and optimizing quantum circuits. 

Figure~\ref{fig:zx} demonstrates a concrete example of the transformation and optimization process. 
In this example, we consider a bell state preparation circuit with four qubits. 
The first step involves transforming the original circuit into a ZX graph representation, under the rules provided by ZX-calculus. 
Once the quantum circuit has been represented as a ZX graph, we can employ optimization methods to simplify the circuit. 
As shown in Figure~\ref{circuitb}, the nodes in the blue boxes of the same color are commutative and can be aggregated into a single node.
We iteratively search for such patterns in the graph, ultimately resulting in a simplified and shallower quantum circuit.
The optimized ZX graph can then be transformed back into a quantum circuit representation, and the circuit depth is reduced from 23 to 14, resulting in a more efficient and compact version of the original bell state preparation circuit. 
This example highlights the power of using ZX-calculus and graph-based optimization techniques to simplify quantum circuits.
Furthermore, we remark that ZX-calculus for optimization is generally universal for most quantum circuits.
For 34 randomly selected circuits, an average depth reduction of 1.48 times is observed as indicated in Figure~\ref{zx_data}.
On extreme occasions, the circuit depth of a variational quantum eigensolver (VQE) is reduced from 7656 to 1110.

\begin{algorithm}[htb]
\caption{Greedy Circuit Partition}
\begin{algorithmic}[1]  
\Procedure{Partition}{$circuit$, $limit$}
\State $blocks \gets \{\}$
\State $groups \gets $ \Call{GroupQubits}{$circuit$}
\For{$group \in groups$}
\State $block \gets $ \Call{CreateBlock}{$group$}
\While{$size(block) < limit$}
\State $gate \gets $ \Call{NextGate}{$circuit$, $block$}
\State \Call{AddGate}{$block$, $gate$}
\EndWhile
\State $blocks \gets blocks \cup \{block\}$
\EndFor
\State \Return $blocks$
\EndProcedure
\Procedure{GroupQubits}{$circuit$}
\State $groups \gets \{\}$
\State $qubits \gets $ \Call{GetQubits}{$circuit$}
\While{$qubits \neq \emptyset$}
\State $q \gets $ \Call{Pop}{$qubits$}
\State $group \gets \{q\} \cup $ \Call{Neighbors}{$q$, $circuit$}
\State $qubits \gets qubits \setminus group$
\State $groups \gets groups \cup \{group\}$
\EndWhile
\State \Return $groups$
\EndProcedure
\end{algorithmic}
\end{algorithm}

\begin{figure*}[t]
    \centering
    \includegraphics[width=0.78\textwidth]{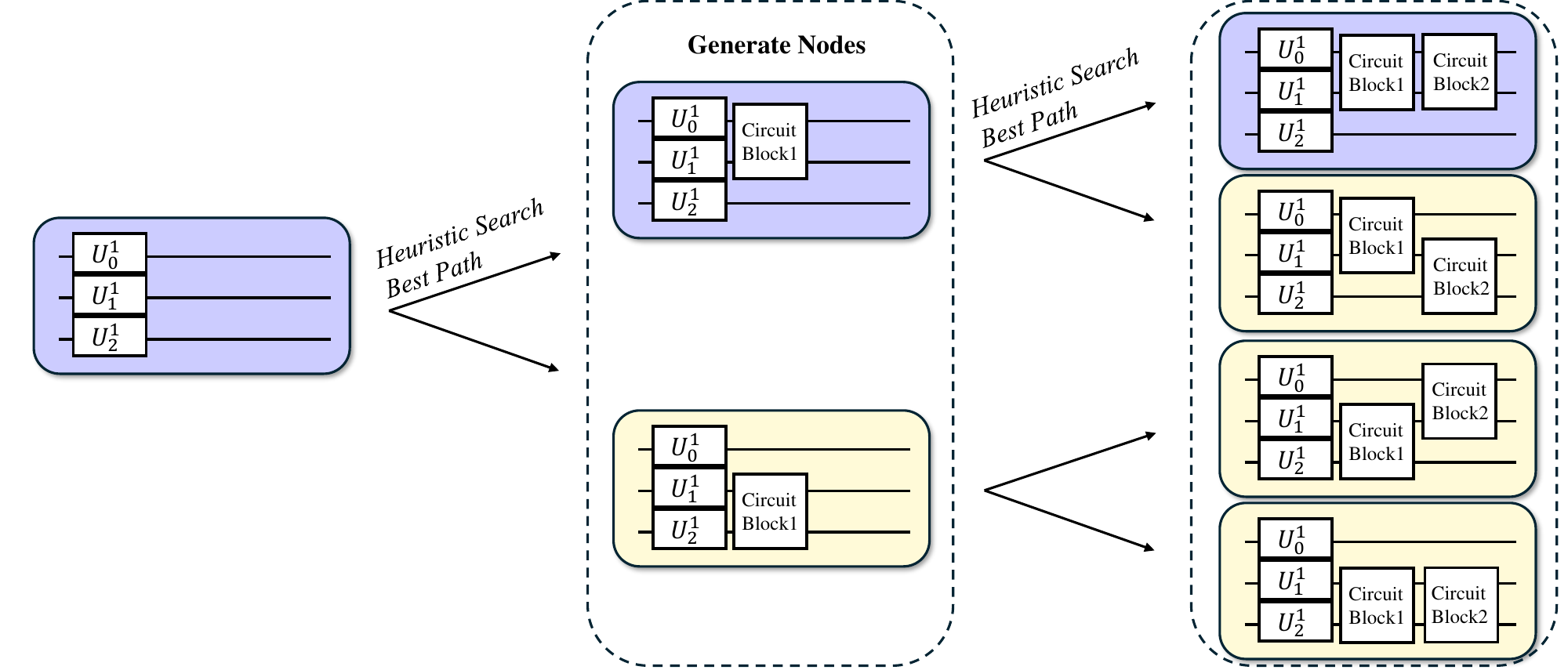}
    \caption{Illustration of the quantum circuit synthesis process using a heuristic search method. 
    The synthesis begins with an empty circuit and iteratively generates new nodes by inserting gates at different positions. 
    Each node represents a candidate quantum circuit. 
    The search for the optimal circuit follows the A\* algorithm, which evaluates nodes based on a cost function that measures the distance between the target unitary matrix and the unitary matrix of the synthesized circuit. 
    The search continues until a circuit that accurately approximates the target unitary matrix with minimal size is found. 
    The resulting synthesized circuit consists of variable unitary gates (VUGs) and CNOT gates.}
    \label{fig:qsearch}
\vspace{-3mm}    
\end{figure*}

\subsection{Greedy Circuit Partition}
\label{partition_section}
After processing the circuit using ZX-calculus, we employ a greedy circuit partitioning algorithm to divide the circuit into smaller circuit blocks, preparing them for synthesis. 
The greedy partitioning algorithm begins by performing horizontal cutting, which involves grouping qubits together to form initial circuit blocks. 
Once the horizontal cutting is complete, we proceed with vertical cutting. 
In this step, we aim to populate each circuit block with as many gates as possible until the block reaches a pre-determined size limit. 
By filling the blocks with gates, we ensure that each block contains a sufficient number of operations to benefit from the subsequent synthesis process. 
The vertical cutting step introduces minimal overhead, allowing us to quickly obtain a list of circuit blocks from the input circuit.
The effectiveness of the greedy partitioning algorithm is enhanced by the prior application of ZX-calculus optimization. 
ZX-calculus simplifies the circuit by reducing it to a small gate set, resulting in a more uniform structure. This uniformity makes the circuit particularly suitable for the greedy partitioning approach, as it allows for a more balanced and efficient distribution of gates among the blocks.
Figure~\ref{circuitc} illustrates the partitioning process, demonstrating a typical circuit block enclosed within a rectangular region after the partitioning step. 
These partitioned circuit blocks are now ready to undergo the subsequent circuit synthesis process, which will further optimize and transform them into a more compact and efficient representation.
By combining ZX-calculus optimization with greedy circuit partitioning, we can effectively decompose large quantum circuits into smaller, more manageable blocks.

\begin{algorithm}[htb]
\caption{Quantum Circuit Synthesis}
\begin{algorithmic}[1]
\Procedure{Synthesize}{$U_{target}$}
    \State $nodes \gets \{\}$
    \State $QC_{empty} \gets $ \Call{CreateEmptyCircuit}{}
    \State $node_{empty} \gets $ \Call{CreateNode}{$QC_{empty}$}
    \State $nodes \gets nodes \cup \{node_{empty}\}$
    \While{$nodes \neq \emptyset$}
        \State $node_{current} \gets $ \Call{SelectNode}{$nodes$}
        \If{\Call{AccuracyThreshold}{$node_{current}, U_{target}$}}
            \State \Return $node_{current}$
        \EndIf
        \State $nodes \gets nodes \setminus \{node_{current}\}$
        \State $successors \gets $ \Call{ExpandNode}{$node_{current}$}
        \For{$successor \in successors$}
            \State $cost \gets $ \Call{CostFunction}{$successor, U_{target}$}
            \State $f \gets cost + $ \Call{Heuristic}{$successor, U_{target}$}
            \State $successor.f \gets f$
            \State $nodes \gets nodes \cup \{successor\}$
        \EndFor
    \EndWhile
\EndProcedure

\Procedure{ExpandNode}{$node$}
    \State $successors \gets \{\}$
    \For{$position \in $ \Call{InsertionPositions}{$node$}}
        \For{$gate \in $ \Call{AvailableGates}{}}
            \State $successor \gets $ \Call{InsertGate}{$node, position, gate$}
            \State $successors \gets successors \cup \{successor\}$
        \EndFor
    \EndFor
    \State \Return $successors$
\EndProcedure
\vspace{-1mm}
\end{algorithmic}
\end{algorithm}
\begin{figure}[t]
\vspace{-3mm}
\centering
   \begin{subfigure}[b]{\linewidth}
   \centering
        \includegraphics[width=0.8\linewidth]{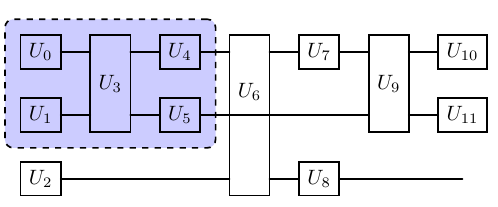}
        \caption{The synthesis result of the partitioned circuit in the blue box of Fig~\ref{circuitc}.}
        \label{synthesisa}
    \end{subfigure}
    \begin{subfigure}[b]{\linewidth}
        \centering
        \includegraphics[width=0.8\linewidth]{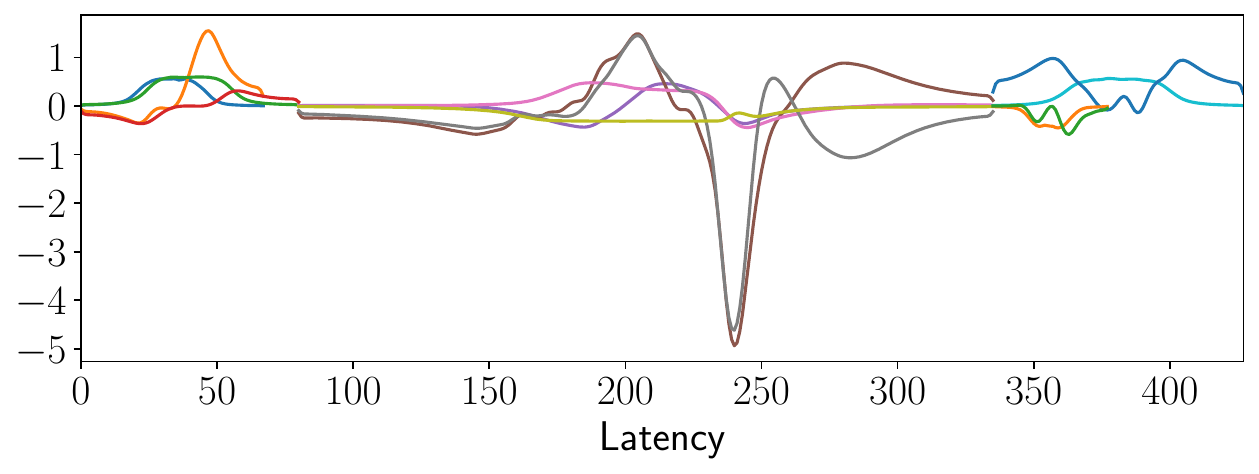}
         \caption{Illustration of the Pulse generation for each gate within the purple box in Fig~\ref{synthesisa}.}
        \label{pulsea}
    \end{subfigure}
    \begin{subfigure}[b]{\linewidth}
        \centering
        \includegraphics[width=0.8\linewidth]{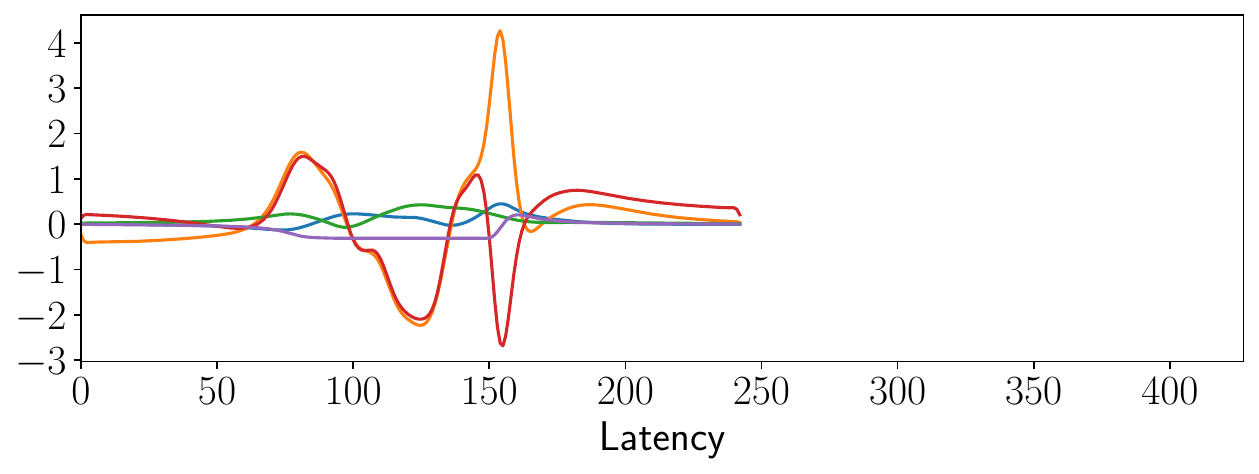}
         \caption{Pulse generation for the entire purple box in Fig~\ref{synthesisa}, treating all gates as a single grouped unitary.}
        \label{pulseb}
    \end{subfigure}
    \caption{Circuit synthesis breaks down large unitary matrices into smaller variable unitary gates (VUGs). The overall pulse latency of the circuit is determined by calculating the latency of each individual VUG. By regrouping these VUGs into larger blocks, the total pulse duration can be significantly reduced.}
    \label{fig:pulse}
\vspace{-3mm}

\end{figure}
\subsection{VUG-based Heuristic Circuit Synthesis}
\label{synthesis_section}

After the greedy circuit partitioning, we obtain a list of grouped gates. 
We then calculate the unitary matrices of these grouped gates. 
Since we limit the size of the groups, the computational overhead here is negligible and totally acceptable. 
In fact, we note that the overhead to compute the unitary matrices remains manageable even with very deep circuit blocks, as long as the number of qubits is moderate.
Once we have the unitary matrices of the circuit blocks, we employ the modified QSearch~\cite{9259942} method within the BQSKIT~\cite{osti_1785933} framework to perform VUG-based circuit synthesis. 
As shown in the Figure~\ref{fig:qsearch}, the synthesis process is based on a heuristic search, where the cost function is defined as the distance between the target unitary and the synthesized circuit. 
The circuit is initialized from scratch, and we use templates to add gates to the circuit and evaluate the distance as the cost function. 
By adding diffrent gates at different positions, we generate various nodes, each representing a quantum circuit. 
We then perform the heuristic search method to find the best node.
During this process, we add variable unitary gates (VUGs) to the circuit template. 
After the synthesis is complete, we obtain a quantum circuit that consists solely of VUGs and CNOT gates. 
Figure~\ref{synthesisa} exhibits the synthesized circuit of the block from Figure~\ref{circuitc}, leading to a depth reduction from 11 to 7.
It is important to note that these VUG gates can be directly used as inputs to the QOC process. 
However, since VUGs are very fine-grained and usually have small sizes, we cannot significantly benefit from QOC when using these VUGs directly.
To address this limitation and fully take advantages of the potential of QOC, we introduce a second regrouping step. 
The purpose of this regrouping is to form slightly larger unitary matrices that can serve as more effective inputs to the QOC process.
As shown in Figure~\ref{pulsea} and \ref{pulseb}, generating pulse directly from VUGs tend to introduce a larger circuit latency.
In contrast, after regrouping the synthesized VUGs in the purple box (Figure~\ref{pulseb}), the pulse latency is largely shortened.

\begin{figure*}
\vspace{-3mm}
    \centering
    \includegraphics[width=0.85\textwidth]{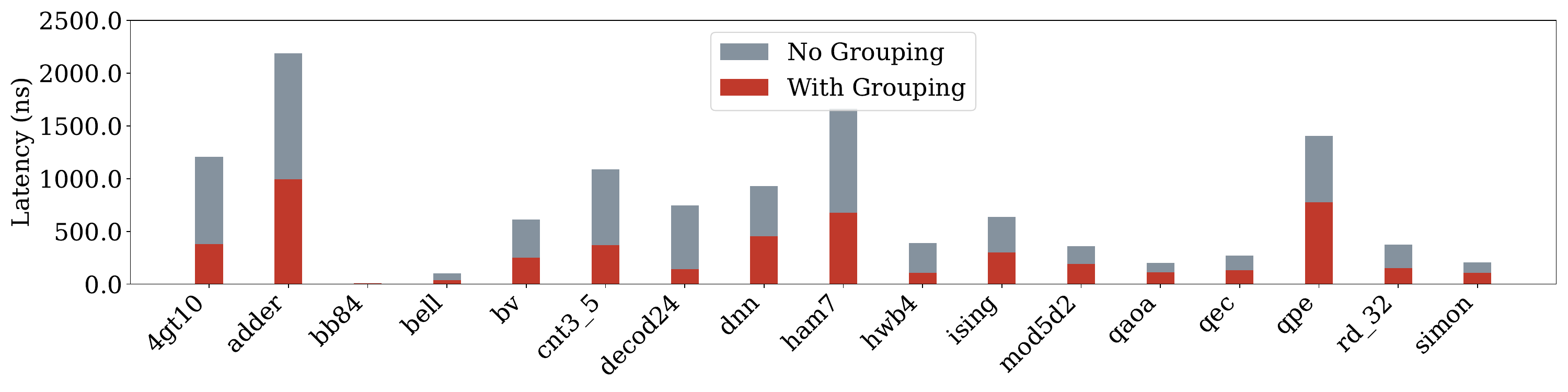}
    \caption{The latency before and after grouping. We show through the results that the grouping can reduce the circuit latency. And in all of our benchmakrs, the grouping latency is shorter than the latency without grouping. This is expected, because we will take advantages of QOC only when the input unitary is not too small to be optimized.}
    \label{fig:latency_comparison-label}
    \vspace{-3mm}
\end{figure*}

\begin{figure*}
    \centering
    \includegraphics[width=0.85\textwidth]{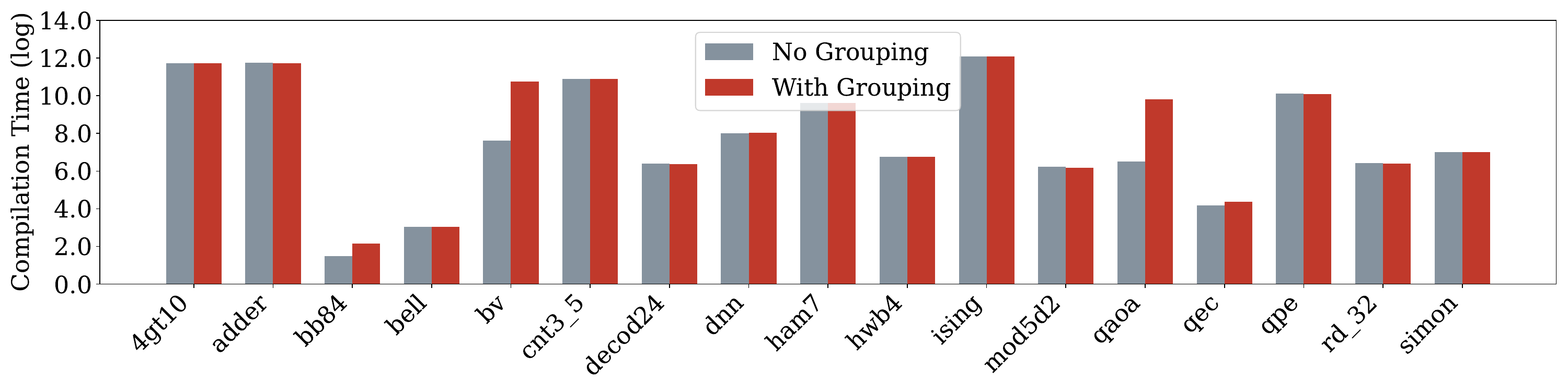}
    \caption{The figure compares the compilation time with and without grouping of quantum gates before the final quantum optimal control step. The results demonstrate that the grouping strategy introduces minimal overhead, and for most of the benchmarks tested, the compilation times are very similar between the two approaches. This indicates that the latency reduction benefits can be achieved through grouping without incurring a significant increase in compilation time.}
    \label{fig:compilation_time_comparison-label}
    \vspace{-3mm}
    
\end{figure*}

\begin{figure*}
    \centering
    \includegraphics[width=0.85\textwidth]{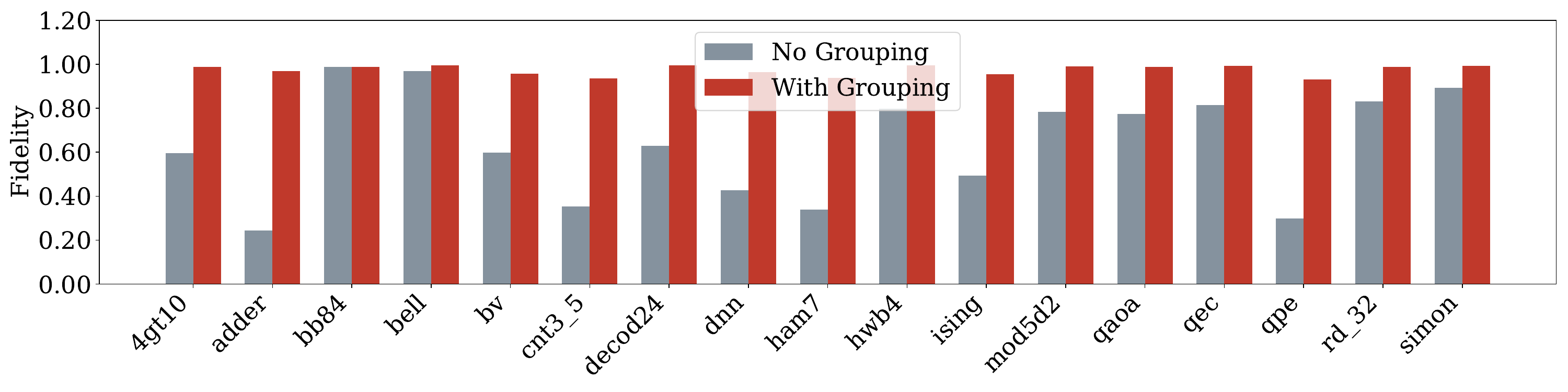}
    \caption{The figure compares the fidelity of quantum circuits with and without grouping. The results show that the fidelities achieved with grouping are generally higher than those without grouping. This can be attributed to the fact that the no-grouping approach operates at a very fine granularity, which can lead to the accumulation of errors during the QOC process. In contrast, the grouping method creates larger unitary matrices and reduces the total number of unitary matrices subjected to QOC. As a result, the grouping approach maintains higher fidelity by mitigating the accumulation of errors.}
    \label{fig:fidelity_comparison-label}
    \vspace{-3mm}
    
\end{figure*}

\subsection{Pulse Generation}
\label{pulse_section}

In our proposed framework \name, our methodologies are built upon prior works AccQOC~\cite{cheng2020accqoc} and PAQOC~\cite{chen2023pulse}. 
We employ QOC techniques to generate a pulse library, which essentially serves as a lookup table. 
The indices of the table are unitary matrices, and each entry stores the corresponding microwave pulses. 
The techniques in AccQOC and PAQOC involve a binary search to determine a short pulse latency for a given unitary matrix as input. 
Specifically, the QOC we employ is implemented using the GRAPE algorithm~\cite{khaneja2005optimal}. 
In the GRAPE algorithm, the pulses are divided into different segments or time slots, with each time slot acting as a unit of time. 
During each time slot, the pulses are assumed to have a constant value. 
It's important to note that the pulses mentioned here refer to the envelope of the final pulses. 
The GRAPE method aims to optimize the time slots according to a cost function, which is often defined as the fidelity between the pulse unitary and the target unitary. 
Based on the cost function and its gradients, the pulse strength during each time slot is adjusted. 
However, pulse simulation and optimization are computationally expensive, which limits the scalability of QOC.
This is the reason why we limit the size of the unitary matrices that serve as inputs to the QOC algorithm.
Once we have obtained the pulse library, we can leverage it to accelerate future QOC processes. 
If we encounter a unitary matrix that is already present in the library, we can directly fetch the pre-existing pulses. 
During this process, we also consider the global phase when comparing unitary matrices. 
In contrast to the previous two works on accelerating QOC, \name\ supports the detection of unitary similarity with global phase. 
By allowing global phase, we can identify more matched unitary matrices, similar to having a higher cache hit rate.

\section{Evaluation}
\label{evaluation}
The experiments were conducted on a Linux-based system with Debian 5.10.179-1 operating system, using 8 nodes. Each node is equipped with 32 cores and 256GB of RAM.

As the first to introduce synthesis into the workflow of quantum optimal control, we compare the results from two settings: (1) applying quantum optimal control directly on the synthesized circuits and (2) grouping the synthesized results before running QOC. The results, shown in Figure~\ref{fig:latency_comparison-label}, Figure~\ref{fig:compilation_time_comparison-label} and Figure~\ref{fig:fidelity_comparison-label}, demonstrate that although both settings have similar compilation times in most scenarios, our method with an extra grouping step significantly reduces circuit latency and increases circuit fidelity. We evaluate our methods using 17 benchmarks from QASMBench~\cite{li2020qasmbench}, showcasing the effectiveness of incorporating the grouping step in the quantum optimal control workflow.

The fidelity is calculated as the accumulated fidelity of each quantum pulse. Specifically, we calculate the fidelity of each pulse and then multiply these fidelities together for all the pulses in the circuit. The fidelity of each individual pulse is determined by the distance between the unitary matrices before and after quantum optimal control. This calculation can be described by the following equation:
\begin{equation}
ESP = \prod_{i=1...n} (1 - |U_i - H_i(t)|)
\end{equation}
where $ESP$ represents the estimated total fidelity of the quantum circuit, $n$ is the number of pulses, $U_i$ is the target unitary matrix for the $i$-th pulse, and $H_i(t)$ is the unitary matrix achieved by the optimized pulse after applying quantum optimal control. The term $|U_i - H_i(t)|$ calculates the distance between the target and optimized unitary matrices for each pulse, and $(1 - |U_i - H_i(t)|)$ represents the fidelity of the $i$-th pulse. The product of these pulse fidelities gives the overall fidelity of the entire quantum circuit.

Across the 17 benchmarks from QASMBench~\cite{li2020qasmbench}, our method achieves an average reduction of 51.11\% in pulse latency with only a 7.11\% increase in compilation time by incorporating a regrouping step after synthesis in the quantum optimal control workflow. Furthermore, our approach manages to increase the circuit fidelity by an average of 33.77\%. These results demonstrate that the regrouping step is necessary and beneficial for reducing circuit latency without introducing significant overhead, highlighting the effectiveness of our proposed method in enhancing the performance of quantum circuits.

We also compare our results with PAQOC~\cite{chen2023pulse}, a recent work on quantum optimal control. As shown in the table~\ref{tabresult}, our approach achieves an average latency reduction of 31.74\% compared to PAQOC. Moreover, we report higher fidelity values for the circuits where data points are available. Although we were unable to obtain some data points from PAQOC, the available data clearly demonstrates that our framework outperforms PAQOC. We attribute this improvement to our novel circuit synthesis process, which effectively optimizes the quantum circuits before the quantum optimal control step, resulting in reduced latency and increased fidelity. 
Synthesis adds extra computational workload compared to PAQOC. However, our synthesis method mainly involves local entanglement and unitary calculations, which can be executed in parallel. Therefore, our approach maintains scalability while reducing pulse latency. We validated our framework by testing it with a large and deep 160-qubit quantum program, obtaining meaningful results. Although this data point isn't in our table due to the lack of comparable PAQOC data, it confirms the feasibility of our method.

\begin{table}
    \centering
    \resizebox{0.49\textwidth}{!}{%
    \begin{tabular}{cccccc}
    \toprule
     \multirow{2}*{Circuit}&   \multicolumn{3}{c}{Latency (ns) } &\multicolumn{2}{c}{Fidelity }  \\
           &  Gate-based & PAQOC~\cite{chen2023pulse} & \name   & PAQOC~\cite{chen2023pulse} & \name \\
           \midrule
        simon   & 469   & 141.23    & \textbf{92}  & -   & \textbf{0.984}\\
        bb84    & 56.5  & 13        & \textbf{10}  & 0.981   & \textbf{0.988}\\
        bv      & 901   & 321       & \textbf{268.5}  & 0.971 &0.968\\
        qaoa    & 1324.5 & 393      & \textbf{111.5}    & 0.952 & \textbf{0.984} \\
        decod24 & 1315.5 & 315      & \textbf{144}      & 0.982 & \textbf{0.989} \\
        dnn     & 3174.5 & 385      & 453.5             & -   & \textbf{0.965}\\
        ham7    & 5238.5 & 1186.5   & \textbf{675.5}    & -   & \textbf{0.938} \\

    \bottomrule
    \end{tabular}
    }
    \caption{Comparison between PAQOC~\cite{chen2023pulse} and \name\ (our approach) for 7 widely applied quantum circuits. In most scenarios, \name\ returns a generation result with shortened latency and increased fidelity. On average, our approach achieves a 31.74\% reduction in latency compared to PAQOC~\cite{chen2023pulse} and a 76.80\% reduction compared with the gate-based method to create pulse sequences. The `-' indicates PAQOC returns fidelity as 0 for the corresponding task.}
    \label{tabresult}
    \vspace{-3mm}
\end{table}

\section{Related Work}
\label{relatedwork}
\textbf{Pulse Generation Framework:} Traditional gate-level circuits in quantum computing have been well-studied, with a wide variety of methods developed for optimizing and mapping qubits at this level~\cite{jin2023exploiting,li2019tackling,zulehner2018efficient}. Recently, researchers have begun to explore a deeper abstraction layer, focusing on pulse-level optimizations to achieve more refined control and efficiency. One previous focus is QOC, QOC shapes external controls on qubits to execute specific tasks. As noted in~\cite{cheng2020accqoc}, QOC can manage quantum circuits of moderate size, although scalability remains a concern. Despite numerous attempts~\cite{sivak2021model,bukov2018reinforcement} to optimize QOC and lower its computational overhead, it continues to be a resource-intensive process.

A recent work AccQOC~\cite{cheng2020accqoc} employs a novel approach by segmenting the quantum circuit into small, uniform subcircuits, each comprising two qubits. It then builds a pulse database for these subcircuits, storing previously generated pulses for efficient reuse. To streamline this process, AccQOC utilizes a similarity graph that maps the distances between different fixed-size subcircuits, including those not currently in the database. The Minimum Spanning Tree (MST) of this graph is then calculated to optimize the order in which control pulses are constructed for the subcircuits. This method significantly enhances the efficiency of pulse compilation, reducing overall time requirements. A more recent work, PAQOC~\cite{chen2023pulse}, is automatically detecting common gate patterns through subgraph mining for expanded search space exploration. It also systemically constructs customized gate-sets based on their impact on overall program latency and quickly adapts to system recalibrations with its small-scale pattern-based gate generation. However, all the aforementioned compilation strategies generate pulses directly from the partitioned circuit group. Therefore, the group size is constrained due to the unmanageable overhead of pulse generation for large unitaries.

\textbf{Synthesis Technologies:} In recent years, to further optimize quantum circuits, multiple methods have been proposed, including Arrays, Decision Diagrams, Tensor Networks, and ZX-Calculus, among others~\cite{wille2022basis}. These approaches each offer unique strengths in addressing the complexities of quantum circuit synthesis and optimization. 
\cite{kissinger2019reducing} have developed a ZX-calculus-based method to reduce non-Clifford T-gates in quantum circuits by up to 50\% without changing circuit structure. This approach uses phase teleportation to cancel non-Clifford phases non-locally and maintains the circuits' functional equivalence. \cite{liu2023tackling} propose a hierarchical qubit mapping and routing algorithm, enhancing quantum circuit efficiency by decomposing circuits into uniformly-sized blocks. Using Permutation-Aware Synthesis (PAS) and Permutation-Aware Mapping (PAM), this method optimizes and maps blocks more effectively than traditional approaches that tend to preserve the original circuit structure.
Compare to these latest advancements, our \name\ framework significantly diminishing circuit durations and elevating pulse performance through innovative applications of bridge synthesis, ZX-Calculus, circuit partitioning, and QOC. This holistic approach not only streamlines quantum computations but also optimizes them to unprecedented levels of efficiency and efficacy.

\section{Conclusion}
\label{conclusion}

In this paper, we introduced the \name\ framework, an advanced integration of ZX-calculus, circuit partitioning, and synthesis techniques with QOC to enhance pulse generation efficiency in quantum circuits. \name\ utilizes a refined approach to granular optimization, significantly reducing quantum circuit latency and computational overhead while increasing parallelism. By strategically regrouping unitary matrices and optimizing their configurations, \name\ not only improves the scalability of QOC processes but also demonstrates considerable performance enhancements in NISQ devices. This methodology effectively balances computational cost with quantum computational power, achieving notable improvements in both pulse generation speed and circuit performance. We achieve an average 31.74\% reduction in circuit latency over PAQOC. Our results indicate a significant potential for applying \name\ in practical quantum computing scenarios, offering a robust solution to the challenges of current quantum technology implementations.

\clearpage
\bibliographystyle{ACM-Reference-Format}
\bibliography{sample-base}

\appendix









\end{document}